\documentclass[11pt]{article}
\usepackage{amsfonts}

\def\1{\'{\i}}
\def\>#1{{\bf #1}}
\def\k{\omega}

\def\R{{\mathbb R}}

\def\adl{L}
\def\ttt{T}
\def\radio{R}

\def\gadl{\hat l}
\def\gtt{\hat t}

\def\dd{{\rm d}}

\def\bbib{\begin{thebibliography}}
\def\ebib{\end{thebibliography}}

\def\be{\begin{equation}}
\def\ee{\end{equation}}
\def\bea{\begin{eqnarray}}
\def\eea{\end{eqnarray}}
\def\disty{\displaystyle}

\newcommand{\nc}{\newcommand}
\nc{\slacs}[1]{\setlength{\arraycolsep}{#1}}
\nc{\ba}{\begin{array}}
\nc{\ea}{\end{array}}
\nc{\I}{{\rm i}}
\nc{\E}{{\rm e}}
\nc{\sfrac}[2]{\mbox{$\frac{#1}{#2}$}}
\nc{\bmth}[1]{\mbox{\boldmath$#1$}}

\begin{document}

 \ \hfil
 
\begin{center}
{\Large{\bf{Quantum  (anti)de Sitter algebras and}}}
  
{\Large{\bf{generalizations of the
kappa--Minkowski space}}}
 
\bigskip

{\Large{\bf{}}}

\end{center}

\bigskip

\begin{center}  Angel Ballesteros, Francisco~J.~Herranz\footnote{Communication presented
in the  XI International Conference on Symmetry Methods in Physics, June 21-24, 2004,
Prague, Czech Republic.\\
Electronically  published in ``Symmetry methods in Physics", edited by  C.
Burdik, O. Navr\'atil and S. Posta, Joint Institute for Nuclear Research, 
Dubna (Russia), pp.  1-20, (2004).}  
\smallskip

  {\it { 
Departamento de F\1sica, Universidad de Burgos, Avda. Cantabria s.n., \\ 09006
Burgos, Spain }}\\ angelb@ubu.es, fjherranz@ubu.es
\end{center}

  \begin{center} 
N.~Rossano Bruno 
 \smallskip

{\it { 
Dipartimento di Fisica, Universit\`a di  Roma Tre  and
INFN Sez.\ Roma Tre,\\ Via Vasca Navale 84, 00146 Roma, Italy}}\\
rossano@fis.uniroma3.it
\end{center}

\bigskip\bigskip

\begin{abstract} 
\noindent
We present two different quantum deformations for the (anti)de
Sitter algebras and groups. The former is a non-standard
(triangular) deformation of $SO(4,2)$ realized as the conformal
group of the (3+1)D Minkowskian spacetime, while the latter is a
standard (quasitriangular) deformation of both $SO(2,2)$ and
$SO(3,1)$ expressed as the kinematical groups of the (2+1)D
anti-de Sitter and de Sitter spacetimes, respectively. The Hopf
structure of the quantum algebra and a study of the dual quantum
group are presented for each   deformation. These results enable
us to propose new non-commutative spacetimes that can be
interpreted as  generalizations of  the $\kappa$--Minkowski space,
either by considering a  variable deformation parameter (depending
on the boost coordinates)   in the conformal deformation, or  by
introducing  an explicit curvature/cosmological constant in the
kinematical one; $\kappa$--Minkowski   turns out to be   the
common first--order structure for all of these quantum spaces.
Some properties provided by these deformations, such as dimensions
of the deformation parameter (related with the Planck length),
space isotropy, deformed boost transformations, etc., are also
commented.
 \end{abstract}

\newpage


\section{Introduction}

Quantum groups have been applied, from the beginning of their
development, in the   construction of deformed   symmetries of
spacetimes~\cite{LukierskiRuegg1992,CGh1,Ita,BCGpho,Giller,Lukierskib,CK3,CK4,Null}
that generalize classical  kinematics beyond   Lie algebras. The
deformation deepest studied is    the well known
$\kappa$--Poincar\'e \cite{LukierskiRuegg1992,Giller,Lukierskib}
which, more recently,     has been applied in the construction of
the so called ``doubly special relativity" (DSR) theories
theories~\cite{Amelino-Camelia:2000mrr,Amelino-Camelia:2000mn,
MagueijoSmolin,Kowalski-Glikman:2002we,Lukierski:2002df} that make
use of two fundamental scales. One is   the usual
observer--independent velocity scale $c$, while the other is an
observer--independent length scale $l_p$ (Planck length) which is
assumed to be related with the deformation parameter. In this way,
DSR theories have established a relationship   between quantum
groups and quantum gravity~\cite{amel,KowalskiFS}.

From the dual quantum group, when the non-commutative spacetime
coordinates $\hat x^\mu$ conjugated to the  $\kappa$--Poincar\'e
translations (momenta) are considered, the        non-commutative
$\kappa$--Minkowski spacetime is
found~\cite{Maslanka,Majid:1994cy,Zak,LukR,LukNR}: \be [\hat
x^0,\hat x^i]=-\frac 1{\kappa}\, \hat x^i\,,\quad [\hat x^i,\hat
x^j]=0\,. \label{xx} \ee More general structures for   quantum
Minkowskian spacetimes have been proposed to be~\cite{Lukierskid}:
\be [\hat x^\mu, \hat x^\nu]=\frac 1{\kappa} (a^\mu\hat
x^\nu-a^\nu\hat x^\mu)\,, \label{xxx} \ee where $a^\mu$ is a
constant four--vector in the  Minkowskian space.

However, in spite of the great activity followed in this field, as
far as we know, there is no an explicit proposal for a quantum
spacetime with a non-zero cosmological constant, or even, some
structure generalizing (\ref{xxx}). In other words, if Lorentz
symmetry has to be modified at the Planck scale and a non-zero
curvature/cosmological constant seems to be physically relevant,
it is necessary  to study the quantum deformations of the (anti)de
Sitter groups. These deformations may provide some deformed
relativistic symmetries, for which the deformed Poincar\'e ones
should be recovered through a flat limit/contraction or zero
cosmological constant.

The aim of this contribution is  to present an overview of some
recent results concerning quantum  (anti)de Sitter algebras and
their dual quantum groups from two different deformations; these
moreover lead to generalized $\kappa$--Minkowski spaces. The
structure of the contribution has two different parts, each of
them deals with   one specific deformation.

In section 2, we present a non-standard quantum deformation of
$so(4,2)$ written as the conformal algebra of the (3+1)D
Minkowskian spacetime~\cite{Herranz:2002fe,Brunoc}. The
Weyl--Poincar\'e algebra (isometries plus dilations) remains as a
Hopf subalgebra after deformation, and this structure is then used
to obtain a quantum   group which, in turn, provides a
non-commutative Minkowskian spacetime. Such a structure involves
the quantum boost parameters in the commutation rules;
alternatively this can also be interpreted as generalization of
(\ref{xx}) with a variable deformation parameter.

In section 3 we study a standard deformation (of Drinfeld--Jimbo
type) for $so(2,2)$ and $so(3,1)$ realized as kinematical algebras
of the (2+1)D anti-de Sitter (AdS) and de Sitter (dS)
spacetimes~\cite{BBH}. We remark that we deal with 2+1 dimensions
as these deformations are not completely constructed yet  in the
proper (3+1)D case. We introduce the quantum algebras and
construct the quantum group through a Weyl quantization of a
Poisson--Lie structure on the (anti)de Sitter groups. When the
non-commutative (anti)de Sitter spacetimes are obtained, it is
shown that they can be interpreted as a generalization of
$\kappa$--Minkowski with a non-zero constant curvature (or
cosmological constant).

For both types of deformations  the deformation parameter is shown
to be related with the Planck length and isotropy of the space is
preserved. Some comments on Lorenzt invariance and deformed boost
transformations are also given.

\section{A non-standard deformation of \bmth{SO(4,2)} in conformal\\ basis}

The Lie algebra $so(4,2)$ of the group of conformal
transformations of the (3+1)D Minkows\-kian  spacetime $\> M^{3+1}$
is spanned by  the generators of rotations $J_i$,  time $P_0$ and
space $P_i$ translations,  boosts $K_i$,  special conformal
transformations $C_\mu$  and  dilations $D$. The non-vanishing
commutation relations of $so(4,2)$ are given by \be
\begin{array}{lll}
[J_i,J_j]=\varepsilon_{ijk}J_k\,,&\quad
[J_i,K_j]=\varepsilon_{ijk}K_k\,,&\quad
[J_i,P_j]=\varepsilon_{ijk}P_k\,,\\[2pt]
[J_i,C_j]=\varepsilon_{ijk}C_k\,,&\quad
[K_i,K_j]=-\varepsilon_{ijk}J_k\,,&\quad
[K_i,P_i]=P_0\,,\\[2pt]
[K_i,P_0]=P_i\,,&\quad  [K_i,C_0]=C_i\,,&\quad
[K_i,C_i]=C_0\,,\\[2pt]
[P_0,C_0]=-2 D\,,&\quad [P_0,C_i]=2K_i\,,&\quad
[C_0,P_i]=2 K_i\,,\\[2pt]
[P_i,C_j]=2(\delta_{ij}D-\varepsilon_{ijk}J_k)\,,&\quad
[D,P_\mu]=P_\mu\,,&\quad [D,C_\mu]=-C_\mu\,,
\end{array}
\label{aa} \ee where throughout this section we  will assume sum
over repeated indices, latin indices $i,j,k=1,2,3$, greek indices
$\mu,\nu=0,1,2,3$, and a system of units such that $c=\hbar=1$; a
generator with three components will be denoted
$\>X=(X_1,X_2,X_3)$.

As is well known, $so(4,2)$ has two remarkable Lie subalgebras:

\noindent $\bullet$  $\{\>J,\>K,\>P,P_0 \}$ that generate the
Poincar\'e subalgebra $\cal P$, that is, the algebra of isometries
of the spacetime $\> M^{3+1}$.

\noindent $\bullet$ $\{\>J,\>K,\>P,P_0,D\}$ that span the
Weyl--Poincar\'e subalgebra  ${\cal WP}$ which is the similitude
algebra of $\> M^{3+1}$.

Hence we have the Lie algebra embedding ${\cal P}\subset{\cal
WP}\subset so(4,2)$.

Alternatively, $SO(4,2)$ can also be interpreted as the  kinematical
group of the (4+1)D AdS  spacetime ${\bf AdS}^{4+1}$. Explicitly
let us  denote by $\adl_{AB}$ $(A<B)$ and $\ttt_{A}$
($A,B=0,1\dots,4$) the Lorentz and translations generators
satisfying \be
\begin{array}{l}
[\adl_{AB},\adl_{CD}]=\eta_{AC}\adl_{BD}-\eta_{AD}\adl_{BC}-\eta_{BC}\adl_{AD}
+\eta_{BD}\adl_{AC}\,,\\[4pt]
\disty
[\adl_{AB},\ttt_{C}]=\eta_{AC}\ttt_{B}-\eta_{BC}\ttt_{A}\,,\qquad
[\ttt_A,\ttt_B]=-\frac 1{\radio^2} \adl_{AB}\,,
\end{array}
\label{aab} \ee such that
$\eta=(\eta_{AB})=\mathrm{diag}\,(-1,1,1,1,1)$ is the Lorentz
metric associated to $so(4,1)$, $\adl_{0B}$ are the four  boosts
in ${\bf AdS}^{4+1}$ and $R$ is the AdS radius related with the
cosmological constant by $\Lambda=6/R^2$. Then  the   change of
basis defined by ($i=1,2,3$): \be
\begin{array}{lll}
\disty \ttt_0=-\frac1{2\radio}\,(C_0+P_0)\,,&\quad
\disty\ttt_1=\frac1{\radio}\,D\,,&\quad
\disty\ttt_{i+1}=\frac1{2\radio}\,(C_i+P_i)\,,\\[9pt]
\disty\adl_{01}=\frac12\,(C_0-P_0)\,,&\quad
\disty\adl_{0,i+1}=K_{i}\,,&\quad
\disty\adl_{1,i+1}=\frac12\,(C_{i}-P_i)\,,\\[9pt]
\disty\adl_{23}=J_3\,,&\quad \disty\adl_{24}=-J_2\,,&\quad
\disty\adl_{34}=J_1\,,
\end{array}
\label{aac} \ee identify the commutation relations (\ref{aa}) with
(\ref{aab}).

\subsection{Conformal Lie bialgebra}

Now we proceed to introduce a quantum deformation of $so(4,2)$ in
the conformal basis (\ref{aa}). The cornerstone of  our
construction is the non-standard or triangular Hopf  algebra that
deforms the Lie algebra with generators $J_3$, $J_+$ verifying \be
[J_3,J_+]= J_+\,, \label{ab} \ee
 and
with classical $r$--matrix~\cite{Drinfelda,Drinfeldb} (fulfilling
the classical Yang--Baxter equation~\cite{Dr}) \be r=z J_3\wedge
J_+ =z(J_3\otimes J_+ - J_+\otimes J_3)\,, \label{ac} \ee where
$z$ is the deformation parameter. The  corresponding deformed
commutator, coproduct and universal quantum $R$--matrix are given
by: \bea
[J_3,J_+]&=&\frac{\E^{z J_+}-1}{z}\,,\nonumber \\
\Delta(J_+)&=&1\otimes J_+ + J_+ \otimes 1\,,\label{ad}\\
\Delta(J_3)&=&1\otimes J_3 + J_3 \otimes \E^{z J_+}\,, \nonumber
\eea \be {\cal R}=\exp\{-z J_+\otimes J_3\}\exp\{z J_3\otimes
J_+\}\,. \label{ae} \ee This structure is, in fact, a Hopf
subalgebra  of many known non-standard quantum  deformations that
cover: $sl(2,\R)\simeq
so(2,1)$~\cite{Demidov,Zakr,Ohn,nonsb,Ogi,nonsd,nonsc,Abdesselam};
$iso(1,1)$, $gl(2)$ and $h_4$~\cite{boson}; the Schr\"odinger
algebra~\cite{schrod}; as well as $so(2,2)$, $so(3,1)$ and
$iso(2,1)$~\cite{vulpiB}. The quantum algebra (\ref{ad}) also
underlies the approach to physics at the Planck scale early
introduced in~\cite{Majida, Majidb}. We  recall that the quantum
algebra (\ref{ad}) is also known as the Jordanian deformation
which is supported by  a twist operator~\cite{Ogi} which relates
the (classical) cocommutative coproduct with the (deformed)
non-cocommutative one while keeping non-deformed commutation
rules~\cite{Drinfeldb}.

The remarkable point is that all of the aforementioned quantum
algebras share the same formal classical $r$--matrix (\ref{ac}),
Hopf subalgebra (\ref{ad}), twisting element and universal
$R$--matrix (\ref{ae}). In the following we proceed to construct a
quantum $so(4,2)$ algebra  starting  again from (\ref{ac}) and
(\ref{ad}).

Let us consider the non-standard classical $r$--matrix (\ref{ac})
written in the conformal basis through the identification \be
D\equiv J_3\,,\quad P_0\equiv J_+\,,\quad \tau\equiv -z\,,
\label{ba} \ee where $\tau$ is now the deformation parameter
related to the usual $\kappa$ and $q$ deformation  parameters
through $\tau=1/\kappa=\ln q$. Next if we assume that \be
r=-\tau D\wedge P_0\,, \label{bb} \ee is  the classical
$r$--matrix for the whole $so(4,2)$ algebra (\ref{aa}), the
corresponding cocommutator $\delta$ of a generic generator $Y_i$
(that defines the associated Lie bialgebra) is obtained  as
$\delta(Y_i)=[ 1\otimes Y_i+Y_i\otimes 1,r]$, namely, \be
\begin{array}{ll}
\delta(P_0)=0\,,\quad &\delta(P_i)=\tau P_i\wedge P_0\,,\\[4pt]
\delta(J_i)=0\,,\quad &\delta(D)=-\tau D\wedge P_0\,,\\[4pt]
\delta(K_i)=-\tau D\wedge P_i\,,\quad &\delta(C_0)=-\tau C_0\wedge
P_0\,,\\[4pt]
\delta(C_i)=-\tau C_i\wedge P_0+2\tau D\wedge K_i\,.\quad&
\end{array}
\label{bc} \ee

The Lie bialgebra $(so(4,2),\delta(r))$ is then formed by the
commutation rules (\ref{aa}) and cocommutators (\ref{bc}) and
determines the structure of both the quantum algebra and its dual
quantum group at the first--order in the deformation parameter,
generators and their dual non-commutative group coordinates. In
other words, the cocommutator (\ref{bc}) characterizes  the
first--order quantum group by means of the  Lie bialgebra duality
\cite{Majidb,CP}, that is, \be \delta(Y_i)=f_i^{jk}Y_j\wedge Y_k
\;\Rightarrow\;[\hat y^j,\hat y^k]=f_i^{jk}\hat y^i\,, \label{bd}
\ee where  $\hat y^i$ is the    quantum group coordinate dual to
$Y_i$    such that $\langle\hat y^i|Y_j\rangle=\delta_j^i$. The
resulting relations provide the underlying first--order
non-commutative spacetime by considering the commutation rules
involving the quantum coordinates dual to the translation
generators (momenta).

In our case, we denote by $\{\hat
x^\mu,\hat\theta^i,\hat\xi^i,\hat d,\hat c^\mu\}$ the dual
non-commutative coordinates of the generators
$\{P_\mu,J_i,K_i,D,C_\mu\}$, respectively. Hence from (\ref{bc})
we obtain the following  non-vanishing first--order quantum group
commutation rules: \be
\begin{array}{lll}
[\hat x^0,\hat x^i]=-\tau \hat x^i\,,&\quad [\hat x^0,\hat d]=\tau
\hat d\,,&\quad
[\hat x^0,\hat c^\mu]=\tau \hat c^\mu\,,\\[4pt]
[\hat d,\hat x^i]=-\tau \hat \xi^i\,,&\quad [\hat d,\hat\xi^i]=2
\tau \hat c^i\,. &
\end{array}
\label{be} \ee Therefore the first--order  quantum Minkowskian
spacetime, ${\bf M}^{3+1}_\tau$, is  given by \be [\hat x^0,\hat
x^i]=-\tau \hat x^i\,,\quad [\hat x^i,\hat x^j]=0\,, \label{bf}
\ee which coincides exactly with the $\kappa$--Minkowski space
(\ref{xx}) provided that $\tau=1/\kappa$. Nevertheless, we shall
compute in  section 2.3 below the  full (all orders) dual quantum
group, thus showing in  section 2.4 that the   complete
non-commutative spacetime generalizes  the $\kappa$--Minkowski
space.

On the other hand, the dual map to (\ref{aac}), which relates
(\ref{be}) with a first--order quantum (4+1)D AdS group in the
kinematical basis, is given by \be
\begin{array}{lll}
\gtt^0=- R (\hat c^0+\hat x^0)\,,&\quad \gtt^1=R\hat d\,,&\quad
\gtt^{i+1}=R(\hat c^{i}+\hat x^{i})\,,\\[4pt]
\gadl^{01}=\hat c^0-\hat x^0\,,&\quad
\gadl^{0,i+1}=\hat\xi^{i}\,,&\quad \gadl^{1,i+1}=\hat c^{i}-\hat x^{i}\,,\\[4pt]
\gadl^{23}=\hat\theta^3\,,&\quad \gadl^{24}=-\hat\theta^2\,,&\quad
\gadl^{34}=\hat \theta^1\,,
\end{array}
\label{bg} \ee where ${\gadl}^{AB}$ and ${\gtt}^A$ are, in this
order, the non-commutative Lorentz and spacetime coordinates dual
to $\adl_{AB}$ and $ \ttt_A$. Hence  the  first--order
non-vanishing  commutation rules for the non-commutative AdS
spacetime, ${\bf AdS}^{4+1}_\tau$, turn out to be \be
[\gtt^0,\gtt^1]=-\tau R\,\gtt^1\,,\quad [\gtt^0,\gtt^{i+1}]=-\tau
R^2 \gadl^{1,i+1}\,,\quad [\gtt^1,\gtt^{i+1}]=-\tau R^2
\gadl^{0,i+1}\,.
 \label{bbgg}
\ee

Therefore, the maps (\ref{aac}) and (\ref{bg}) (or some kind of
non-linear generalization if higher orders in $\tau$ and in the
coordinates were considered) would  allow one to express the same
quantum deformation of $so(4,2)$  within two physically different
frameworks, thus relating  deformed conformal Minkowskian and
kinematical AdS symmetries. Such a quantum group relationship
might  further be applied in order to analyze the role that
quantum deformations of $so(4,2)$ could play in relation  with the
``AdS--CFT correspondence" that relates local QFT on ${\bf
AdS}^{d+1}$ with a conformal QFT on the  compactified
Minkowskian spacetime ${\bf CM}^{(d-1)+1}$~\cite{Maldacena,
Witten,Reh}. We remark that  the connection for $d=3$
corresponding to a three--parameter quantum $o(3,2)$ algebra has
been studied in~\cite{Mozb}.

\subsection{Quantum conformal algebra}

The Hopf structure  of the quantum $so(4,2)$ algebra,
$U_\tau(so(4,2))$, can be obtained by applying a direct
construction~\cite{vulpiB}. Firstly, we deduce the coproduct
$\Delta$  as follows.

\noindent
\begin{itemize}

\item[$\bullet$]  Require that (\ref{ad}) remains as a Hopf subalgebra of
$U_\tau(so(4,2))$.

\item[$\bullet$] Take into account that the cocommutator $\delta$ (\ref{bc})
corresponds to the skewsymmetric part of the first--order
deformation of the complete coproduct $\Delta$: \be
\Delta=\sum_{k=0}^{\infty} \Delta_{(k)}=\sum_{k=0}^{\infty} \tau^k
\delta_{(k)}\,, \quad   \delta =\delta_{(1)} - \sigma\circ
\delta_{(1)}\,, \label{ca} \ee where  $\sigma(X\otimes Y)=Y\otimes
X$.

\item[$\bullet$] And solve the coassociativity condition
$(1\otimes \Delta)\Delta=(\Delta\otimes 1)\Delta$.

\end{itemize}

The resulting coproduct turns out to be~\cite{Herranz:2002fe} \be
\begin{array}{rcll}
\Delta(P_0)&=&1\otimes P_0 + P_0\otimes 1\,,\\[4pt]
\Delta(P_i)&=&1\otimes P_i + P_i\otimes\E^{\tau P_0}\,,\\[4pt]
\Delta(J_i)&=&1\otimes J_i + J_i\otimes 1\,,\\[4pt]
\Delta(K_i)&=&1\otimes K_i + K_i\otimes 1-\tau D\otimes\E^{-\tau P_0}P_i\,,\\[4pt]
\Delta(D)&=&1\otimes D+D\otimes\E^{-\tau P_0}\,,\\[4pt]
\Delta(C_0)&=&1\otimes C_0+C_0\otimes\E^{-\tau P_0}\,,\\[4pt]
\Delta(C_i)&=&1\otimes C_i+C_i\otimes\E^{-\tau
P_0}+2\tau D\otimes\E^{-\tau P_0}K_i-\\[4pt]
&&-\tau^2(D^2+ D)\otimes\E^{-2\tau P_0} P_i\,.
\end{array}
\label{cc} \ee

Secondly, the deformed commutation rules are deduced by imposing
$\Delta$ to be an algebra homomorphism, that is,
$\Delta([X,Y])=[\Delta(X),\Delta(Y)]$; these are written in two
sets~\cite{Herranz:2002fe}:

\noindent $\bullet$ Commutation relations which  close a
Weyl--Poincar\'e Hopf subalgebra $U_\tau({\cal WP})\subset \\
U_\tau(so(4,2))$: \be
\begin{array}{lll}
[J_i,J_j]=\varepsilon_{ijk}J_k\,,&\quad
[J_i,K_j]=\varepsilon_{ijk}K_k\,,&\quad [J_i,P_0]=0\,,\\[2pt]
[J_i,P_j]=\varepsilon_{ijk}P_k\,,&\quad
[K_i,K_j]=-\varepsilon_{ijk}J_k\,,&\quad [P_\mu,P_\nu]=0\,,\\[4pt]
[K_i,P_0]=\E^{-\tau P_0}P_i\,,&\quad \disty
[K_i,P_j]=\delta_{ij}\,\frac{\E^{\tau P_0}-1}{\tau}\,,&\quad
[D,K_i]=0\,,\\[6pt]
[D,P_i]=P_i\,,&\quad \disty [D,P_0]=\frac{1-\E^{-\tau
P_0}}{\tau}\,,&\quad [D,J_i]=0\,.
\end{array}
\label{cd}
\ee

\noindent $\bullet$ Commutation relations  that involve  the
special conformal transformations $C_\mu$: \be
\begin{array}{ll}
[J_i,C_j]=\varepsilon_{ijk}C_k\,,&\quad [J_i,C_0]=0\,,\\[4pt]
[C_i,C_j]=0\,,&\quad [C_0,C_i]=-\tau(D C_i + C_i D)\,,\\[4pt]
[K_i,C_0]=C_i\,,&\quad [K_i,C_j]=\delta_{ij}(C_0 -\tau D^2)\,,\\[4pt]
[P_i,C_j]=2\delta_{ij}D-2\varepsilon_{ijk}J_k\,,&\quad
[C_0,P_i]=2 K_i+\tau(D P_i+ P_iD)\,,\\[4pt]
[P_0,C_0]=-2D\,,&\quad [P_0,C_i]=\E^{-\tau P_0}K_i+K_i\E^{-\tau
P_0}\,,\\[4pt]
[D,C_i]=-C_i\,,&\quad [D,C_0]=-C_0+\tau D^2\,.
\end{array}
\label{ce}
\ee

Finally, the counit and antipode maps can directly be derived from
the Hopf algebra axioms and we omit them.

By construction, some relevant Lie subalgebras of $so(4,2)$ are
promoted to Hopf subalgebras of $U_\tau(so(4,2))$ after
deformation. In particular, we find the following $so(p,q)$ and
Weyl--Poincar\'e Hopf subalgebras, all of them containing the
generators $P_0$ and $D$, and sharing the  same classical
$r$--matrix (\ref{bb}): \be
\begin{array}{ll}
U_\tau(sl(2,\R))\simeq U_\tau(so(2,1))&\qquad \{D,P_0,C_0\}\cr
\qquad \phantom{\cap}\qquad \qquad \qquad \cap&\qquad\qquad  \cap\cr
U_\tau({\cal
{WP}}^{1+1}) \subset  U_\tau(so(2,2))&\qquad
 \{D,P_0,P_1,K_1;\ C_0,C_1\}\cr
\qquad \cap\qquad \qquad \qquad \cap&\qquad\qquad  \cap\cr
U_\tau({\cal
{WP}}^{2+1}) \subset  U_\tau(so(3,2)) &\qquad  \{D,P_0,P_1,P_2,K_1,K_2,J_3;\
C_0,C_1,C_2\} \cr
\qquad \cap\qquad \qquad \qquad \cap&\qquad\qquad  \cap\cr
U_\tau({\cal
{WP}}^{3+1}) \subset   U_\tau(so(4,2))&\qquad \{D,P_0,\>K,\>J;\ C_0,\>C\}
 \end{array}
\label{cf} \ee However, the Poincar\'e subalgebras do  not remain
as Hopf subalgebras after this deformation; this is a consequence
of the   presence of  the dilation generator in the coproduct of
the boosts (\ref{cc}). A similar  fact was also pointed out for
some (standard) Drinfeld--Jimbo deformations in \cite{Vlado}.

The chain of embeddings (\ref{cf})  ensures that  properties
previously known for a given low dimensional deformation can
directly be extended to higher dimensional deformations. In this
respect, let us consider the universal $R$--matrix (\ref{ae})
written in the conformal basis by applying the map (\ref{ba}): \be
{\cal  R}=\exp\{ \tau P_0\otimes D\}\exp\{-\tau D\otimes P_0\}\,.
\label{cg} \ee This element has been shown to be a universal
$R$--matrix for $U_\tau(sl(2,\R))$~\cite{nonsd}, that is, this
fulfils the quantum Yang--Baxter equation and the property \be
{\cal  R}\Delta(X){\cal  R}^{-1}=\sigma\circ \Delta(X)\quad
\mbox{for}\quad X\in\{D,P_0,C_0\}\,. \label{ch} \ee For the
remaining generators of $U_\tau(so(4,2))$,
$X\in\{\>J,\>P,\>K,\>C\}$, it can be proven that \be
\begin{array}{l}
\exp\{-\tau D\otimes P_0\}\Delta(X)\exp\{\tau D\otimes P_0\} =
1\otimes X+X\otimes 1 \equiv \Delta_0(X)\,,\\[4pt]
\exp\{ \tau P_0\otimes D\}\Delta_0(X)\exp\{ -\tau P_0\otimes D\} =
\sigma\circ  \Delta(X)\,,
\end{array}
\label{ci} \ee so that the element  (\ref{cg})  is also the
universal $R$--matrix for $ U_\tau(so(4,2))$ as well as for
\textit{all} the quantum algebras arising in the embeddings
(\ref{cf}).

Another application conveyed by (\ref{cf}) is the extension of the
time discretization of the (1+1)D massless Klein--Gordon (or wave)
equation~\cite{vulpiB} associated to $U_\tau(so(2,2))$ to (2+1)D
with $U_\tau(so(3,2))$~\cite{Czech} and  to (3+1)D with
$U_\tau(so(4,2))$~\cite{Herranz:2002fe}. The generators of all of
these quantum algebras have been realized as
differential--difference operators acting on a uniform Minkowskian
spacetime lattice discretized along the time direction (the space
coordinates remain continuous) and with the deformation parameter
$\tau$ playing the role of  the time lattice constant.

\subsection{Quantum Weyl\bmth{-}Poincar\'e group}

The existence of the universal $R$--matrix (\ref{cg}) enables, in
principle,  to deduce the  quantum group dual to any of the
quantum algebras appearing in (\ref{cf}) by applying the
Faddeev--Reshetikhin--Takhtajan (FRT) procedure~\cite{Faddeev:ih}.
Such an approach requires a matrix representation of the chosen
quantum algebra as well as a matrix element $\hat{\cal T}$ of the
quantum group with non-commutative entries.

In what follows we shall restrict ourselves to deal with the
quantum group dual to $U_\tau({\cal WP})$  instead of that dual to
the complete $U_\tau(so(4,2))$ since for the latter it is not
possible to identify properly the quantum space and time
coordinates but only formal non-commutative matrix entries.

The change of basis (\ref{aac}) allows us to deduce a $6\times6$
deformed matrix representation of $U_\tau(so(4,2))$ in the
conformal basis (fulfilling (\ref{cd}) and (\ref{ce})) by starting
from the vector representation of the (4+1)D quantum AdS algebra;
namely \be
\begin{array}{l}
\disty
P_0=\frac{\tau}{2}\,\bigl(e_{00}- e_{01}+ e_{10}-
e_{11} \bigr)-e_{02}-e_{12}+e_{20}-e_{21}\,,\\[4pt]
P_i=e_{0,i+2}+e_{1,i+2}+e_{i+2,0}-e_{i+2,1}\,,\qquad
D= e_{01}+e_{10}\,,\\[4pt]
J_i=- \varepsilon_{ijk}e_{j+2,k+2}\,,\qquad
K_i=e_{2,i+2}+e_{i+2,2}\,, \\[4pt]
C_0=\tau \bigl(e_{00}+ e_{11}\bigr)-e_{02}+e_{12}+e_{20}+e_{21}\,,\\[4pt]
C_i=e_{0,i+2}-e_{1,i+2}+e_{i+2,0}+e_{i+2,1}\,,
\end{array}
\label{db} \ee where $e_{ab}$ ($a,b=0,\dots,5$) is the $6\times6$
matrix with entries $(e_{ab})_{ij}=\delta_{ai}\delta_{bj}$. Hence
a $6\times6$ matrix representation for the quantum
Weyl--Poincar\'e algebra (\ref{cd}) arises within  (\ref{db}).

Under this representation $P_0^3$ vanishes, so that the  quantum
${R}$--matrix (\ref{cg}) reduces to the $36\times 36$ matrix given
by \be {\cal R}=\left(\mbox{\boldmath $1$}\otimes \mbox{\boldmath
$1$}+\tau P_0\otimes D +\sfrac12\,\tau^2 P_0^2\otimes D^2 \right)
\left( \mbox{\boldmath $1$}\otimes \mbox{\boldmath $1$}-\tau
D\otimes P_0 +\sfrac12\,\tau^2 D^2\otimes P_0^2 \right),
\label{df} \ee where $\mbox{\boldmath $1$}$ is the $6\times 6$
unit matrix.

Next we construct the quantum Weyl--Poincar\'e group element
$\hat{\cal T}$ by considering the following matrix product that
depends on the matrix generators (\ref{db}) and their dual
non-commutative coordinates: \bea \hat{\cal T}&=&{\rm e}^{\hat d
D} \,{\rm e}^{\hat x^0 P_0} \left( {\rm e}^{\hat x^1 P_1} \,{\rm
e}^{\hat x^2 P_2} \,{\rm e}^{\hat x^3 P_3} \right) \left( {\rm
e}^{\hat \theta^1 J_1} \,{\rm e}^{\hat \theta^2 J_2} \,{\rm
e}^{\hat \theta^3 J_3}\right) \left( {\rm e}^{\hat \xi^1 K_1}
\,{\rm e}^{\hat \xi^2 K_2} \,{\rm e}^{\hat \xi^3
K_3}\right)=\nonumber\\&=&{\slacs{1ex}\left(\begin{array}{cccccc}
\hat\alpha_+&\hat\beta_-&\hat\gamma_0&
\hat\gamma_1&\hat\gamma_2&\hat\gamma_3\\
\hat\beta_+&\hat\alpha_-&\hat\gamma_0&
\hat\gamma_1&\hat\gamma_2&\hat\gamma_3\\
\hat x^0&- \hat x^0&\hat\Lambda^0_0&
\hat\Lambda^0_1&\hat\Lambda^0_2&\hat\Lambda^0_3\\
\hat x^1&- \hat x^1&\hat\Lambda^1_0&
\hat\Lambda^1_1&\hat\Lambda^1_2&\hat\Lambda^1_3\\
\hat x^2&- \hat x^2&\hat\Lambda^2_0&
\hat\Lambda^2_1&\hat\Lambda^2_2&\hat\Lambda^2_3\\
\hat x^3&- \hat x^3&\hat\Lambda^3_0&
\hat\Lambda^3_1&\hat\Lambda^3_2&\hat\Lambda^3_3
\end{array}\right).}
\label{dd} \eea The non-commutative entries are just the quantum
Minkowskian coordinates $\hat x^\mu$, the formal   Lorentz entries
$ \hat\Lambda^\mu_{\nu} =\hat\Lambda^\mu_{\nu}(\hat \theta^i,\hat
\xi^i),$ which involve quantum rotation  and boost coordinates,
verifying \be \hat\Lambda^\mu_{\nu} \hat\Lambda^\rho_{\sigma}
g^{\nu\sigma} =g^{\mu\rho}\,,\quad  \hat x_\mu=g_{\mu\nu} \hat
x^\nu\,,\quad (g^{\mu\rho})=\mathrm{diag}\,(-1,1,1,1)\,,
\label{de} \ee as well some functions   $\hat \alpha$, $\hat\beta$
and $\hat\gamma$ of   the quantum coordinates which are defined by
\be
\begin{array}{l}
\hat\alpha_\pm=\cosh\hat d\pm \frac 12\,{\rm e}^{\hat d}\bigl(\hat
x_\mu\hat x^\mu +\tau\hat x^0\bigr)\,,\quad \hat\gamma_\nu={\rm
e}^{\hat d} \hat x_\mu
\hat \Lambda^\mu_\nu\,,\\[4pt]
\hat\beta_\pm=\sinh\hat d\pm \frac 12\,{\rm e}^{\hat d}\bigr(\hat
x_\mu\hat x^\mu +\tau\hat x^0\bigl)\,. \ea \label{dde} \ee

Notice that if the complete quantum $SO(4,2)$ group were
considered by adding the remaining exponentials of the conformal
transformations (\ref{db}) to the product (\ref{dd}), the
non-commutative coordinates $\hat x^\mu$ will no longer appear as
themselves, thus precluding a further and direct identification of
the associated non-commutative spacetime as usually happens when
dealing with quantum deformations of semisimple groups (see,
e.g.,~\cite{Chang} for the construction of a standard
$q$--$SO(3,2)$).

Now the FRT approach gives rise to the commutation rules,
coproduct, counit and antipode by means of the relations \be {\cal
R}\hat{\cal T}_1\hat{\cal T}_2=\hat{\cal T}_2\hat{\cal T}_1{\cal
R}\,,\quad \Delta(\hat{\cal T})=\hat{\cal T}\dot\otimes \hat{\cal
T}\,,\quad \epsilon(\hat{\cal T})=\>1\,,\quad   S(\hat{\cal
T})=\hat{\cal T}^{-1}\,, \label{dg} \ee respectively, and where
$\hat{\cal T}_1=\hat{\cal T}\otimes \mbox{\boldmath $1$}$ and
$\hat{\cal T}_2=\mbox{\boldmath $1$}\otimes\hat{\cal T}$. The
resulting commutation rules and coproduct initially depend on all
the entries of $\hat{\cal T}$, but they can be further and
consistently reduced, with the aid of (\ref{dde}), to expressions
that only depend on $\{\hat d,\hat x^\mu, \hat\Lambda^\mu_{\nu}
\}$; these are~\cite{Brunoc} \be
\begin{array}{l}
\Delta (\hat x^\mu)=\hat x^\mu \otimes {\rm e}^{-\hat d}+
\hat \Lambda^\mu_{\eta}\otimes \hat x^\eta\,,\\[4pt]
\Delta(\hat d)=\hat d\otimes 1+ 1\otimes \hat d\,,\\[4pt]
\Delta({\hat \Lambda^\mu_{\nu}})= \hat \Lambda^\mu_{\eta} \otimes
\hat \Lambda^\eta_{\nu}\,, \label{di}
\end{array}
\ee \be
\begin{array}{l}
[\hat d,\hat\Lambda^\mu_{\nu} ]=0\,,\qquad
 [\hat x^\alpha,\hat\Lambda^\mu_{\nu}]=0\,,\qquad
[\hat \Lambda^\alpha_{\beta},\hat\Lambda^\mu_{\nu} ]=0\,,\\[4pt]
 [\hat d,\hat x^\mu]=\tau\left(
\delta^\mu_{0}{\rm e}^{-\hat d}-\hat \Lambda^\mu_{0} \right),\qquad
[\hat x^\mu , \hat x^\nu ] =\tau
  \left(\hat \Lambda^\nu_{0}\hat x^\mu -
\hat \Lambda^\mu_{0} \hat x^\nu   \right),
\end{array}
\label{dh} \ee where the quantum Lorentz entries
$\hat\Lambda^\mu_{0}$  are given by \be
\begin{array}{ll}
\hat \Lambda^0_0=\cosh\hat\xi^1\cosh\hat\xi^2\cosh\hat\xi^3\,,\quad&
\hat\Lambda^2_0= \sinh\hat\xi^2\cosh\hat\xi^3\,,\\[4pt]
\hat\Lambda^1_0=\sinh\hat\xi^1\cosh\hat\xi^2\cosh\hat\xi^3\,,\quad&
\hat\Lambda^3_0= \sinh\hat\xi^3\,.
\end{array}
\label{dj}
\ee

The commutation relations (\ref{dh}) show that the functions
$\hat\Lambda^\mu_{\nu}$ are indeed commuting quantities, so that
there are no ordering problems in any of the above expressions;
this, in turn, implies that $[\hat\xi^i,\hat\xi^j]=0$.

Notice also that if we take  in (\ref{dh}) the first--order in all
the quantum coordinates (in this case   $\hat\Lambda^0_0\to 1$ and
$\hat\Lambda^i_0\to \hat\xi^i$),  we recover  the relations
defining the Weyl--Poincar\'e bialgebra in its dual form as \be
[\hat x^0,\hat x^i]=-\tau \hat x^i\,,\quad [\hat d,\hat x^0]=-\tau
\hat d\,,\quad [\hat d,\hat x^i]=-\tau \hat \xi^i\,,
 \label{dk}
\ee
which  coincide with (\ref{be}) provided that $\hat c^\mu\equiv 0$.

\subsection{Non-commutative Minkowskian spacetime}

The non-commutative Minkowskian spacetime  ${\bf M}_\tau^{3+1}$
with quantum Weyl--Poincar\'e group symmetry is identified within
the set of commutation rules (\ref{dh}) by considering  those
involving the quantum coordinates $\hat x^\mu$~\cite{Brunoc}: \be
[\hat x^\mu , \hat x^\nu ] =\tau
  \left(\hat \Lambda^\nu_{0}(\hat\xi) \hat x^\mu -
\hat \Lambda^\mu_{0}(\hat\xi) \hat x^\nu   \right) . \label{dl}
\ee This  can be interpreted as a  generalization of (\ref{xxx})
through the map  $a^\mu\to\hat\Lambda^\mu_{0}(\hat\xi)$, where the
Lorentz entries involved  only depend on the quantum boost
coordinates as given in (\ref{dj}).

The commutativity character of $\hat \Lambda^\mu_{0}(\hat\xi)$
shown in (\ref{dh}) suggests that these can be regarded as
structure constants in   (\ref{dl}). As a byproduct,    the
quantum boost coordinates $\hat\xi^i$  are   commutative
coordinates (so scalars) within the quantum Weyl--Poincar\'e
group. We would also like to mentioning that if the quantum
conformal transformations $C_\mu$ and parameters $\hat c^\mu$ were
taken into account  and the corresponding quantum $SO(4,2)$ group
were constructed, then $\hat\xi^i$ and $\hat \Lambda^\mu_{0}$   no
longer would commute with the dilation parameter $\hat d$, as
follows from  the first--order relations (\ref{be}), in such a
manner that (\ref{dl}) would define a quadratic non-commutative
Minkowskian spacetime.

An alternative form to express (\ref{dl}),  formally closer to
$\kappa$--Minkowski, is achieved by introducing new quantum space
coordinates $\hat X^i$  defined by \be \hat x^0 \rightarrow \hat
x^0\,,\quad \hat x^i \rightarrow   \hat X^i= \hat x^i\hat
\Lambda^0_0(\hat\xi)-\hat x^0\hat\Lambda^i_0(\hat\xi)\,.
\label{dm} \ee The transformed ${\bf M}_\tau^{3+1}$  reads \be
[\hat X^i,\hat x^0]=\tau \hat\Lambda^0_0(\hat\xi)\hat X^i\,,\quad
[\hat X^i,\hat X^j]=0\,,\label{dn} \ee which shows a
generalization of the $\kappa$--Minkowski space (\ref{xx}) with a
``variable" deformation parameter
$\tau'(\hat\xi)=\tau\cosh\hat\xi^1\cosh\hat\xi^2\cosh\hat\xi^3$.

Now we shall comment on some of the  properties derived from the
above results.

\subsubsection{Dimensions of the deformation parameter}

Dimensional analysis  of this deformation (see, e.g.,  expressions
(\ref{bb}) or (\ref{dk}))  shows that  the deformation parameter
$\tau$ is endowed with the same dimensions as the Planck length
$l_P$ (recall that we consider units with $c=\hbar=1$), so inverse
to the parameter $\kappa$; these    are inherited either from
$P_0$ or from $\hat x^0$: \be [\tau]=[P_0]^{-1}=[\hat x^0]\,,\quad
[\tau]=\frac1{[\kappa]}\,. \label{ddnn} \ee Therefore $\tau$ is a
dimensionful deformation parameter that can be  considered to be
related with the Planck length, thus playing the role of an
observer--independent   (fundamental) scale. In this respect, we
remark that $U_\tau({\cal WP})$ has allowed us to construct  a DSR
theory, different from the proposals coming from
$\kappa$--Poincar\'e, which can be found in~\cite{dsrbruno}.

\subsubsection{Space isotropy}

The explicit form of ${\bf M}_\tau^{3+1}$ (\ref{dl}) (also
(\ref{dn})) shows that the quantum rotation coordinates
$\hat\theta^i$ are absent; these only appear as arguments of the
quantum Lorentz entries
$\hat\Lambda^\mu_{i}=\hat\Lambda^\mu_{i}(\hat\theta,\hat\xi)$. In
this sense we can say that the isotropy of the space is preserved.

The same property follows from the Hopf structure of $U_\tau({\cal
WP})$. The coproduct (\ref{cc}) exhibits non-deformed
(cocommutative) rotation generators (this a direct consequence of
the bialgebra (\ref{bc}) since $\delta(J_i)=0$), while the
deformed commutation relations  (\ref{cd}) show that $\>P$ and
$\>K$ are transformed as classical vectors under rotations.  In
fact, $\>J$ close a non-deformed $so(3)$ algebra.

\subsubsection{Lorentz invariance}

The explicit dependence of $\hat\Lambda^\mu_{0}$ on the quantum
boost coordinates $\hat \xi$ in ${\bf M}_\tau^{3+1}$ indicates
that  different observers in relative motion with respect to
quantum group transformations have a different perception of the
spacetime non-commutativity which, in turn, implies that Lorentz
invariance is lost.

Nevertheless, from our point of view, the required property in the
context of quantum groups  should be Lorentz coinvariance rather
than Lorentz invariance. This means that the commutation rules
(\ref{dl})  that define ${\bf M}_\tau^{3+1}$  should be
coinvariant under  quantum group transformations, that is, under
the transformation laws for the quantum coordinates which are
provided by the coproduct (\ref{di}). Covariance of  (\ref{dl})
under such quantum group transformations is ensured by construction and
we refer to~\cite{Brunoc} for more details.

\subsubsection{Boost transformations}

The deformed finite  boost transformations  have been obtained
within the DSR theory developed in~\cite{dsrbruno} by working with
the  quantum algebra $U_\tau({\cal WP})$. Such transformations
close a group as in the non-deformed case and moreover the
additivity of the boost parameter (rapidity) for two deformed
transformations  along a same direction is   preserved. In this
respect, we remark that these properties are in full agreement
with the commutation rule  $[\hat\xi^i,\hat\xi^j]=0$ provided by
the quantum group.

Finally, we also recall that the range of boost parameters, energy
$P_0$ and momenta $\>P$ deeply depend on the  sign of the
deformation parameter $\tau$, so that  two different scenarios
appear~\cite{dsrbruno}.

\section{A standard deformation of \bmth{SO(2,2)} and \bmth{SO(3,1)} in\\
kinematical basis}

The Lie algebras of the isometry groups of the three (2+1)D
relativistic spacetimes of constant curvature  $\k$ can be
described in a unified way by means of the curvature itself, which
plays the role of a  graded contraction parameter~\cite{CK3}; we
denote this family of Lie algebras by $so_\k(2,2)$. If
$\{J,P_0,P_i, K_i\}$   are, in this order, the generators of
rotations, time translations, space translations and boosts,   the
commutation relations of  $so_\k(2,2)$ read \be
\begin{array}{lll}
[J,P_i]=\epsilon_{ij}P_j\,,&\quad
[J,K_i]=\epsilon_{ij}K_j\,,&\quad  [J,P_0]=0\,,\\[4pt]
[P_i,K_j]=-\delta_{ij}P_0\,,&\quad [P_0,K_i]=-P_i\,,&\quad
[K_1,K_2]=-J\,,\\[4pt]
[P_0,P_i]=\k K_i\,,&\quad [P_1,P_2]=-\k J\,,\end{array} \label{ea}
\ee where, in contrast with  section 2, we now assume  that Latin
indices $i,j=1,2$, Greek ones $\mu,\nu=0,1,2$, and $\epsilon_{ij}$
is a skewsymmetric tensor such that $\epsilon_{12}=1$.

According to the sign of  $\k$  we find that the Lie brackets
(\ref{ea}) reproduce:

\noindent
$\bullet$   The  AdS   algebra, $so(2,2)$,
when  $\k=+1/R^2>0$ and  where $R$ is the AdS radius.

\noindent $\bullet$ The   dS   algebra, $so(3,1)$, when
$\k=-1/R^2<0$  and   where $R$ is the dS radius.

\noindent
$\bullet$ And the  Poincar\'e   algebra, $iso(2,1)$, when $\k=0$; this case
also corresponds to the flat limit/contraction $R\to \infty$  such that   $so(2,2)\to
iso(2,1)\leftarrow so(3,1)$.

\subsection{(Anti)de Sitter Lie bialgebras}

The quantum deformation we are going to deal with in this section
is based in the Drinfeld--Jimbo deformation of $sl(2,\R)\simeq
so(2,1)$~\cite{Drinfelda}, $U_z(sl(2,\R))$, whose classical
$r$--matrix, deformed commutation rules and coproduct in the basis
$\{J_3,J_\pm\}$ are given by \be r=z J_+\wedge J_-\,,
\label{eb}\ee \be \label{ec} \ba{l} \disty [J_3,J_\pm]=\pm
J_\pm\,,\qquad
[J_+,J_-]=\frac{\sinh(2zJ_3)}{z}\,,\\[4pt]
\Delta(J_3)=1 \otimes J_3 +J_3\otimes 1\,,\\[4pt]
\Delta (J_\pm)=\E^{- z J_3}\otimes
J_\pm+J_\pm\otimes\E^{zJ_3}\,.\ea\ee The $r$--matrix (\ref{eb}) is
a solution of the modified classical Yang--Baxter equation, so
that $U_z(sl(2,\R))$ is a deformation of standard or
quasitriangular type.

The well known Lie algebra isomorphism $so(2,1)\oplus
so(2,1)\simeq so(2,2)$ can be implemented in the quantum group
framework. As far as the classical $r$--matrix is concerned, the
difference of two copies of (\ref{eb}) with the same deformation
parameter gives rise to a classical $r$--matrix of
$so(2,2)$~\cite{beyond}, here written in the kinematical basis \be
r=z(K_1\wedge P_1+K_2\wedge P_2)\,. \label{ed} \ee This element is
again a solution of the modified classical Yang--Baxter equation,
not only for $so(2,2)$ but also for $so(3,1)$ and $iso(2,1)$, no
matter of the curvature $\k$. In fact, this is exactly the
$r$--matrix underlying  the (2+1)D $\kappa$--Poincar\'e algebra
provided that $z=1/\kappa$. Hence we can take (\ref{ed}) as the
$r$--matrix for the whole family $so_\k(2,2)$.

By following the same procedure of section 2.1 we obtain that the
cocommutator coming from (\ref{ed}) reads \bea
\delta(P_0)&=&0\,, \qquad \delta(J)=0\,,\nonumber\\
\delta(P_i)&=&z(P_i\wedge P_0-\k\epsilon_{ij} K_j\wedge J)\,, \label{ee}\\
\delta(K_i)&=&z(K_i\wedge P_0+\epsilon_{ij} P_j\wedge
J)\,.\nonumber \eea Meanwhile the  dual non-vanishing commutation
rules turn out to be \be
\begin{array}{ll}
[\hat\theta,\hat x^i]= z\epsilon_{ij}  \hat \xi^j\,,&\quad
[\hat x^0, \hat x^i]=-z \hat x^i\,,\\[4pt]
[\hat\theta,\hat \xi^i]=-z\k \epsilon_{ij}  \hat x^j\,,&\quad
[\hat x^0, \hat \xi^i]=-z \hat \xi^i\,,
\end{array}
\label{ef} \ee where $\{\hat\theta,\hat x^\mu,\hat\xi^i\}$ are the
non-commutative group coordinates dual to the generators
$\{J,P_\mu,\\ K_i\}$, respectively.

Consequently,  from (\ref{ef})  we find that the first--order
non-commutative AdS,   Minkows\-kian and dS spacetimes are
simultaneously defined by the (2+1)D $\kappa$--Minkowski space
(similarly to     the previous deformation): \be [\hat x^0,\hat
x^i]=-z \hat x^i\,,\quad [\hat x^1, \hat x^2]=0\,. \label{eg} \ee
As it can be expected, when  higher orders in the quantum
coordinates are considered the resulting non-commutative spaces
generalize the $\kappa$--Minkowski one since corrections depending
on the curvature appear; thus we shall present in section 3.4
below three different quantum spaces, all of them sharing the same
first--order relations (\ref{eg}).

\subsection{Quantum (anti)de Sitter algebras}

Let us consider two copies of the quantum algebra (\ref{ec}) such
that the two deformation parameters only differ by the sign. Then
the direct sum of quantum algebras~\cite{Ita,beyond,CGST2} \be
U_z(so(2,1))\oplus U_{-z}(so(2,1))\simeq U_z(so(2,2))\,, \ee
together with a contraction analysis lead to the Hopf structure of
the standard quantum deformation of (\ref{ea}) and (\ref{ee})
denoted  $U_z(so_\k(2,2))$~\cite{CK3}: \be
\begin{array}{rcl}
\Delta(P_0)&=&1\otimes P_0+P_0\otimes 1\,,\qquad
\Delta(J)=1\otimes J+J\otimes 1\,,\\[6pt]
\Delta(P_i)&=&\E^{-zP_0/2}\cosh(\frac{z}2\,\rho J)\otimes P_i+
P_i\otimes\E^{zP_0/2}\cosh(\frac{z}2\,\rho J)+\\[4pt]
&&+\rho\E^{-zP_0/2}\sinh(\frac{z}2\,\rho J)\otimes\epsilon_{ij}
K_j-\rho\epsilon_{ij}K_j\otimes\E^{zP_0/2}\sinh(\frac{z}2\,\rho J)\,,\\[6pt]
\Delta(K_i)&=&\E^{-zP_0/2}\cosh(\frac{z}2\,\rho J)\otimes K_i+
K_i\otimes\E^{zP_0/2}\cosh(\frac{z}2\,\rho J)-\\[4pt]
&&\disty -\E^{-zP_0/2}\,\frac{\sinh(\frac{z}2\,\rho
J)}{\rho}\otimes\epsilon_{ij}P_j+
\epsilon_{ij}P_j\otimes\E^{zP_0/2}\,\frac{\sinh(\frac{z}2\,\rho
J)}{\rho}\,,
\end{array}
\label{eh}
\ee
  \be
\begin{array}{ll}
[J,P_i]=\epsilon_{ij}P_j\,,\qquad[J,K_i]=\epsilon_{ij}K_j\,,\qquad&[J,P_0]=0\,,\\[4pt]
\disty [P_i,K_j]=-\delta_{ij}\,\frac{\sinh (zP_0)}{z}\,\cosh(z\rho
J)\,,\qquad& [P_0,K_i]=-P_i\,,\\[6pt]
\disty [K_1,K_2]=-\cosh(zP_0)\,\frac{\sinh(z\rho
J)}{z\rho}\,,\qquad&
[P_0,P_i]=\k K_i\,,\\[6pt]
\disty [P_1,P_2]=-\k \cosh(zP_0)\,\frac{\sinh(z\rho
J)}{z\rho}\,,&\end{array} \label{ei} \ee where hereafter we  also
express the curvature as $\k=\rho^2$. 

Therefore $\rho=1/R$ for
$U_z(so(2,2))$, $\rho=\I/R$ for $U_z(so(3,1))$, and the
contraction $\k=0$ to $\kappa$--Poincar\'e $U_z(iso(2,1))$
corresponds to the flat limit $\rho\to 0$. We remark that such a
contraction is always well defined in any of the expressions
presented in this section, so that it is not necessary to perform
any kind of quantum In\"on\"u--Wigner contractions~\cite{LBC} as
we deal with a built-in scheme of contractions~\cite{CK3}. In
particular, the limit $\rho\to 0$ of $U_z(so_\k(2,2))$ directly
gives rise to the coproduct and deformed commutation rules of
$\kappa$--Poincar\'e as \be
\begin{array}{rcl}
\Delta(P_0)&=&1\otimes P_0+P_0\otimes 1\,,\qquad
\Delta(J)=1\otimes J+J\otimes 1\,,\\[4pt]
\Delta(P_i)&=&\E^{-zP_0/2}\otimes P_i+
P_i\otimes\E^{zP_0/2}\,,\\[4pt]
\Delta(K_i)&=&\E^{-zP_0/2}\otimes K_i+
K_i\otimes\E^{zP_0/2}-\\[2pt]
&&\disty -\frac{z}2\,\epsilon_{ij} \left(\E^{-zP_0/2}J\otimes P_j-
P_j\otimes\E^{zP_0/2}J\right),
\end{array}
\label{ej}
\ee
  \be
\begin{array}{l}
[J,P_i]=\epsilon_{ij}P_j\,,\qquad [J,K_i]=\epsilon_{ij}K_j\,,\qquad
[J,P_0]=0\,,\\[4pt]
\disty [P_i,K_j]=-\delta_{ij}\,\frac{\sinh(zP_0)}{z}\,,\qquad
[P_0,K_i]=-P_i\,,\\[4pt]
[K_1,K_2]=-J\cosh(zP_0)\,,\qquad\quad [P_\mu,P_\nu]=0\,.
\end{array}
\label{ek}
\ee

\subsection{Quantum (anti)de Sitter  groups: a Poisson\bmth{-}Lie structure}

As we have already commented in the section 2.3 for
$U_\tau(so(4,2))$, to obtain the complete   quantum group dual to
a quantum deformation of a semisimple Lie algebra is, in general,
a cumbersome task that requires to know a matrix representation of
both the quantum group element $\hat{\cal T}$ and the quantum
$R$--matrix $\cal R$;   the FRT procedure can then be applied.
However,  even if these objects are known, the quantum spacetime
and boost coordinates would appear ``hidden" as arguments of some
formal non-commutative entries of the matrix   $\hat{\cal T}$.

Another way to get an insight into the non-commutative structures
associated to the quantum group is to  compute the  Poisson--Lie
brackets  provided by the classical $r$--matrix for the
commutative coordinates, and next to analyzing their possible
non-commutative version.  The  steps of this procedure are as
follows.

\begin{itemize}

\item[$\bullet$]
Obtain a matrix element $\cal T$ of the Lie group (so with commutative
coordinates) by means of a matrix representation of the algebra.

\item[$\bullet$]
From $\cal T$, compute left $Y_i^L$ and right $Y_i^R$ invariant
vector fields for each algebra generator $Y_i$.

\item[$\bullet$]
Construct the Poisson--Lie structure on the group that comes from
the classical $r$--matrix  $r= r^{ij}Y_i\otimes Y_j$  through the
Sklyanin bracket defined by \cite{Dr}: \be \{f,g\}=
r^{ij}(Y_i^Lf\, Y_j^L g - Y_i^Rf\, Y_j^R g)\,, \label{gb} \ee
where $f,g$ are smooth functions of the Lie   group coordinates
$y^i$. In this way the Poisson--Lie brackets for  $y^i$ can be
obtained, say \be \{y^i,y^j\}=zF(y^k)\,, \label{gc} \ee where
$F(y^k)$  is a smooth function depending on some set of
coordinates $y^k$.

\item[$\bullet$]
Finally, apply the Weyl substitution of the initial  Poisson
brackets between commutative coordinates (\ref{gc}) by commutators
between non-commutative coordinates~\cite{ Dr,Tak}. \be [\hat
y^i,\hat y^j]=z F(\hat y^k)+o(z^2)\,. \label{gd} \ee

\end{itemize}

Obviously, there is no guarantee that the Weyl quantization gives
the complete  quantum group dual to the initial quantum algebra,
specially when dealing with semisimple groups, since ordering
problems often appear during the quantization procedure. However,
this approach provides, at least, the non-commutative structure up
to second--order in the deformation parameter and in \textit{all}
orders in the quantum coordinates; note that the $o(z^2)$ term in
(\ref{gd}) comes from the reordering of the  quantum coordinates
$\hat y^k$ within the function $F$. Recall that the
$\kappa$--Poincar\'e group~\cite{Maslanka,Majid:1994cy,Zak,LukR}
(in any dimension) has been constructed by applying the above
procedure. In fact, for this deformation there does not exist a
universal $R$--matrix except for the (2+1)D
case~\cite{Maslanka,CGST2,karpacz}.

In our case, we start with the following 4D real matrix
representation of $so_\k(2,2)$ (verifying (\ref{ea})): \be
\begin{array}{ll}
P_0=-\k e_{01}+e_{10}\,,&\quad P_i=\k e_{0\,i+1}+e_{i+1,0}\,,\\[4pt]
J=-e_{23}+e_{32}\,,&\quad  K_i=  e_{1\,i+1}+e_{i+1,1}\,,
 \end{array}
\label{ge} \ee where $e_{ab}$ ($a,b=0,\dots,3$) is the $4\times4$
matrix with entries $(e_{ab})_{ij}=\delta_{ai}\delta_{bj}$. Under
this representation, any generator $Y\in so_\k(2,2)$ fulfils \be
Y^{T}\mathbb I +\mathbb I\, Y=0\,,\quad \mathbb
I=\mathrm{diag}\,(1,\k,-\k,-\k)\,, \label{gf} \ee where $Y^T$ is
the transpose matrix of $Y$. Next we construct a $4\times4$ matrix
element of the group $SO_\k(2,2)$, under the representation
(\ref{ge}), through the following product: \be T=\exp(x^0
P_0)\exp(x^1 P_1)\exp(x^2 P_2) \exp(\xi^1 K_1)\exp(\xi^2 K_2)
\exp(\theta J)\,. \label{gg} \ee Left  and right invariant vector
fields, $Y^L$ and $Y^R$, of $SO_\k(2,2)$ and  the Poisson--Lie
structure associated to the $r$--matrix (\ref{ed})  can then be
computed and they can explicitly be found in~\cite{BBH}. We only
present here the Poisson--Lie brackets involving group spacetime
$x^\mu$ and boost $\xi^i$ coordinates: \be
\{x^0,x^1\}=-z\,\frac{\tanh\rho x^1}{\rho \cosh^2\rho x^2}\,,\quad
\{x^0,x^2\}=-z\,\frac{\tanh\rho x^2}{\rho}\,,\quad
\{x^1,x^2\}=0\,, \label{gh} \ee \be
\{\xi^1,\xi^2\}=z\rho\,\frac{\sinh\rho x^1}{\cosh \rho
x^2}\left(\frac{\sinh\rho x^1\tanh\rho
x^2\sinh\xi^1+\cosh\xi^1\sinh\xi^2}{\cosh\rho
x^1}-\frac{\tanh\xi^2}{\cosh\rho x^2}\right), \label{gi} \ee and
the remaining $\{x^i,\xi^j\}\ne 0$ for any value of
$\k\equiv\rho^2$.

We stress that the order of the matrix product (\ref{gg}) is not
arbitrary but this is chosen in such a manner that $x^0$ and $x^i$
correspond, in this order, to geodesic distances measured along
time--like and space--like geodesics~\cite{BBH}.  These quantities
are called ``geodesic parallel coordinates"  and they can be
interpreted as Cartesian coordinates on curved spacetimes. In
particular, the metric  on the (2+1)D spacetimes reads \be
\dd\sigma^2 =\cosh^2(\rho x^1) \cosh^2(\rho x^2)(\dd
x^0)^2-\cosh^2(\rho x^2)(\dd x^1)^2-(\dd x^2)^2\,, \label{ggj} \ee
which under the limit $\rho\to 0$ reduces to the Minkowskian
metric in  flat Cartesian coordinates  $\dd\sigma^2=(\dd
x^0)^2-(\dd x^1)^2-(\dd x^2)^2$.

\subsection{Non-commutative (anti)de Sitter spacetimes}

The Poisson bracket $\{x^1,x^2\}=0$ allows us to propose (but not
to prove!) that the defining commutation relations of the (2+1)D
non-commutative AdS and dS spacetimes are  a direct Weyl
quantization of  the  Poisson--Lie brackets (\ref{gh}), as no
ordering problems appear in the   commutation rules. By expanding
them in power series in the curvature we find that \bea [\hat x^0,
\hat x^1]&=&-z\, \frac{\tanh\rho\hat x^1}{\rho\cosh^2\rho\hat x^2}
=-z\hat x^1+\frac{z}3\,\k(\hat x^1)^3+z\k\hat x^1
(\hat x^2)^2 +o(\k^2)\,,\nonumber\\[1pt]
[\hat x^0,\hat x^2]&=&-z\,\frac{\tanh\rho\hat x^2}{\rho}=
-z\hat x^2+\frac{z}3\,\k(\hat x^2)^3+o(\k^2)\,,\label{ha}\\[1pt]
[\hat x^1,\hat x^2]&=&0\,. \nonumber \eea Hence the linear
relations in $\hat x^i$ correspond to the common ``seed",
$\kappa$--Minkowski, while corrections on the curvature start to
arising in the third--order. Now three different quantum
spacetimes are found; when $\k\ne 0$ these can be interpreted as
generalizations of $\kappa$--Minkowski with a non-zero
cosmological constant. Notice that the ``asymmetric" form of
(\ref{ha}) could be expected from the beginning, as for instance
the classical metric (\ref{ggj}) shows. This is a consequence of
dealing with intrinsic coordinates $x^i$ of the spacetime. However
it is possible to express (\ref{ha}) in a complete symmetric form
by introducing non-commutative analogues of the 4D ambient space
coordinates, where the (2+1)D spaces can be embedded~\cite{BBH}.

To end with, we briefly comment on some properties of this
deformation as well as on some open problems.

\begin{itemize}

\item[$\bullet$]
Similarly to the previous $\tau$ (\ref{ddnn}), the deformation parameter
$z$ is a dimensionful one such that
$[z]=[P_0]^{-1}=[\hat x^0]=1/[\kappa]$,  so that
this can also be related with  the Planck length $l_P$.

\item[$\bullet$]
Space isotropy is preserved at both the quantum algebra and group
levels: $\hat\theta$ is absent from (\ref{ha}), the rotation $J$
remains primitive in  (\ref{eh}), and  $\>P,\>K$ are transformed
as classical vectors under $J$ (\ref{ei}).

\item[$\bullet$]
Since the coproduct for the quantum $SO_\k(2,2)$ group is still
unknown, we have no quantum group transformations, under which,
the quantum spacetimes should be coinvariant.

\item[$\bullet$]
The geometrical/physical role of the quantum coordinates  $\hat
x^i$ deserves a further study; their classical counterpart
suggests that they may be interpreted as some kind of ``geodesic
operators".

\item[$\bullet$]
In $\kappa$--Poincar\'e with $\k=0$, the bracket (\ref{gi})
vanishes, so in the quantum case $[\hat\xi^i,\hat\xi^j]=0$. This
is  consistent with the study developed in
\cite{Bruno:primo,Bruno:2002wc} showing that $\kappa$--Poincar\'e
boost  transformations  close a  group  and also that additivity
of rapidity is preserved. Nevertheless, this is no longer true
when a non-zero curvature is considered as
$[\hat\xi^i,\hat\xi^j]\ne 0$, so that these properties   may be
either lost or somewhat modified for the quantum (anti)de Sitter
algebras.

\end{itemize}

\bigskip\noindent
{\small This work was partially supported  by the Ministerio de
Educaci\'on y Ciencia (Spain, Project FIS2004-07913), by the Junta
de Castilla y Le\'on   (Spain, Project  BU04/03), and by the
INFN-CICyT (Italy--Spain).}

\end{document}